# Prevention of cyberattacks in WSN and packet drop by CI framework and information processing protocol using AI and Big Data

Shreyanth S
Department of Electronics and Communication Engineering,
Anna University, Chennai, India
E-mail: shreyanth0712@gmail.com

*Abstract*—As the reliance on wireless sensor networks (WSNs) rises in numerous sectors, cyberattack prevention and data transmission integrity become essential problems. This study provides a complete framework to handle these difficulties by integrating a cognitive intelligence (CI) framework, an information processing protocol, and sophisticated artificial intelligence (AI) and big data analytics approaches. The CI architecture is intended to improve WSN security by dynamically reacting to an evolving threat scenario. It employs artificial intelligence algorithms to continuously monitor and analyze network behavior, identifying and mitigating any intrusions in real time. Anomaly detection algorithms are also included in the framework to identify packet drop instances caused by attacks or network congestion. To support the CI architecture, an information processing protocol focusing on efficient and secure data transfer within the WSN is introduced. To protect data integrity and prevent unwanted access, this protocol includes encryption and authentication techniques. Furthermore, it enhances the routing process with the use of AI and big data approaches, providing reliable and timely packet delivery. Extensive simulations and tests are carried out to assess the efficiency of the suggested framework. The findings show that it is capable of detecting and preventing several forms of assaults, including as denial-of-service (DoS) attacks, node compromise, and data tampering. Furthermore, the framework is highly resilient to packet drop occurrences, which improves the WSN's overall reliability and performance.

I. INTRODUCTION

Wireless Sensor Networks (WSNs) have become an essential component of a wide range of applications, from environmental monitoring to industrial automation. These networks are made up of many small, low-cost sensor nodes that work together to collect data and send it to a central base station for processing. However, extensive deployment of WSNs brings weaknesses that make them vulnerable to cyberattacks and packet drop occurrences. It is vital to ensure the security and reliability of data transmission in WSNs in order to protect the integrity of gathered data and avoid any interruptions in key systems.

Researchers and practitioners now consider the prevention of intrusions and the mitigation of packet drop incidents in WSNs to be of paramount importance. Traditional security measures and routing protocols are frequently insufficient to address the constantly changing threat landscape and the dynamic nature of WSNs. Consequently, there is a growing need for innovative approaches that can detect and mitigate intrusions proactively while ensuring reliable packet delivery.

This research addresses the growing susceptibility of WSNs to cyberattacks and the ensuing packet loss incidents. Cyberattacks on WSNs can result in a variety of negative outcomes, including illicit access, data manipulation, and disruption of vital services. In addition, packet loss incidents can occur as a result of network intrusions or congestion, jeopardizing the integrity and timeliness of data transmission. Therefore, a robust solution is required to protect WSNs from cyber threats and guarantee uninterrupted data transmission.

It also utilizes a comprehensive architecture to avoid cyberattacks in WSNs and mitigate packet drop problems. A cognitive intelligence (CI) framework, an information processing protocol, and sophisticated techniques based on artificial intelligence (AI) and big data analytics are all part of the proposed framework. The framework intends to improve the security, integrity, and reliability of WSNs by using these technologies, tackling the key concerns related with cyberattacks and packet drop.

There are three research objectives for this paper. First, design and implement a CI framework that continuously monitors and analyses network behaviour in real-time, mitigating potential intrusions. Second, devise a protocol for information processing that ensures secure and efficient data transmission within the WSN, thereby optimizing routing and minimizing packet loss incidents. Third, integrate AI and big data techniques into the framework to improve the overall security, dependability, and robustness of WSNs.

The suggested architecture is built around the CI foundation. It constantly monitors the WSN's behaviour, assessing numerous network characteristics and trends in order to detect potential cyberattacks. The framework makes use of AI methods such as machine learning and pattern recognition to detect unusual activity and conduct suitable countermeasures in real time. Isolating compromised nodes, altering network parameters, or dynamically adjusting security policies are examples of countermeasures (Table I).

TABLE I. FUNCTIONALITIES OF COMPONENTS IN THE PROPOSED FRAMEWORK

| Component | Functionality |
|---|---|
| Cognitive Intelligence | Real-time monitoring and analysis of network behavior to detect and mitigate cyberattacks |
| Framework | Dynamically adapting to the evolving threat landscape and adjusting security policies |
| Information Processing | Encryption and authentication mechanisms to ensure secure and efficient data transmission |
| Protocol | Optimizing routing process using AI and big data techniques for reliable packet delivery |
| AI and Big Data | Analyzing large volumes of data to identify complex attack patterns |
| Techniques | Predicting potential threats and facilitating proactive decision-making |

An information processing protocol is designed to function in tandem with the CI architecture to provide secure and efficient data transmission within the WSN. To ensure the integrity and confidentiality of transmitted data, this protocol includes encryption and authentication techniques. Furthermore, it enhances the routing process with AI and big data techniques, improving packet delivery reliability and timeliness. The protocol avoids congested routes and dynamically adjusts to network conditions by using an adaptive routing technique, reducing the likelihood of packet drop occurrences.

The incorporation of AI and big data approaches within the architecture improves the WSN's overall security and reliability. AI algorithms evaluate massive volumes of WSN data, allowing for the detection of complex attack patterns and the prediction of potential threats. Big data analytics techniques provide important insights into network behaviour and enable proactive decision-making in order to avoid intrusions and mitigate packet loss issues.

## II. LITERATURE REVIEW

Ozdemir and Xiao (2009) provide an exhaustive review of secure data aggregation in wireless sensor networks (WSN). The paper discusses the difficulties associated with data aggregation and proposes several techniques for ensuring secure and reliable data transmission. The authors analyse existing data aggregation protocols and mechanisms and emphasize their strengths and weaknesses. The study provides valuable insights into the field of secure data aggregation in WSNs and serves as a foundational reference for preventing intrusions and packet loss in WSNs [1]. Gomathi and Mahendran (2015) present efficient scheduling schemes for data packets in wireless sensor networks (WSN). The paper introduces novel techniques to optimize the packet scheduling process, with the goal of minimizing packet loss and ensuring packet delivery reliability. The authors evaluate the performance of the proposed schemes and their efficacy in enhancing the data transmission efficiency in WSNs. This research is a valuable resource for enhancing packet scheduling in WSNs and preventing cyberattacks [2]. Dilek, Çakır, and Aydın (2015) provide an exhaustive overview of the applications of artificial intelligence (AI) techniques in the fight against cybercrime. The paper examines the efficacy of various AI-based approaches, such as machine learning, data mining, and expert systems, in detecting and preventing cyberattacks. The authors examine the advantages and disadvantages of these techniques and emphasize their potential to enhance cybersecurity. This article is a useful resource for incorporating AI techniques into the prevention of intrusions in wireless sensor networks and ensuring reliable data transmission [3]. Wallgren, Raza, and Voigt (2013) examine routing attacks and countermeasures in the Internet of Things (IoT) based on RPL. This paper examines the flaws in the RPL routing protocol and outlines a variety of countermeasures against routing attacks. The authors assess the efficacy of the proposed countermeasures and analyse their contribution to enhancing the security of IoT networks. This study contributes significant knowledge to the prevention of intrusions in wireless sensor networks and reinforces the need for robust routing protocols [4]. Wu, Liu, Zhou, and Zhan (2012) propose a three-layer brain-like learning-based intelligent intrusion prediction method for wireless sensor networks (WSNs). This paper presents a novel framework for predicting intrusions in WSNs using artificial intelligence techniques. The authors demonstrate the accuracy with which the three-layer brain-like learning model can detect and predict intrusions. This research contributes valuable insights to the prevention of intrusions in WSNs by highlighting the role of intelligent prediction techniques in enhancing security [5]. Moon, Iqbal, and Bhat (2016) introduce an authenticated key exchange protocol for wireless sensor networks (WSN). This paper presents a novel protocol that assures secure key exchange between sensor nodes, thereby enhancing the network's overall security. The authors assess the performance and efficacy of the proposed protocol and emphasize its capacity to prevent unauthorized access and assaults in WSNs. This research contributes valuable insights to intrusion prevention and the establishment of secure communication in WSNs [6]. Watro et al. (2004) propose TinyPK, a novel public key technology-based method for enhancing the security of sensor networks. This paper presents a lightweight public key infrastructure for sensor nodes with limited resources. The authors demonstrate TinyPK's efficacy in securing communications and preventing unauthorized access. This research contributes significant insights to the prevention of intrusions in wireless sensor networks, highlighting the significance of efficient security mechanisms for devices with limited resources [7].

## III. ARCHITECTURE AND FLOW

The proposed framework for the prevention of intrusions in WSNs and the mitigation of packet drop incidents is intended to provide an all-encompassing solution that guarantees the security, integrity, and dependability of data transmission. The architecture of the framework consists of three essential elements: the Cognitive Intelligence (CI) framework, the information processing protocol, and advanced techniques based on artificial intelligence (AI) and big data analytics. Figure 1 depicts the architecture of the proposed framework, illustrating the data flow and interaction between the various components.

The framework's data flow begins with the sensor nodes of the WSN collecting data from the surrounding environment. This information is then transmitted to the CI framework for monitoring and analysis in real time. Continuously analysing network parameters such as data traffic, node activity, and communication patterns, the CI framework monitors the network's behaviour in real-time. Using AI algorithms such as machine learning and pattern recognition, the CI framework identifies potential intrusions and anomalous behaviour that may result in packet loss incidents.

The CI framework initiates the appropriate mitigation measures upon identifying a cyberattack or prospective threat. These measures may include isolating compromised nodes, reconfiguring network parameters, or dynamically modifying security policies to prevent additional unauthorized access or data manipulation. The CI framework guarantees the WSN's resilience against cyber threats by continuously adapting to the changing threat landscape and deploying proactive defence mechanisms.

Simultaneously, the sensor nodes' collected data is transmitted securely via the information processing protocol. This protocol employs encryption and authentication mechanisms to safeguard the integrity and privacy of transmitted data. Using AI and big data methods, the information processing protocol also optimizes the routing process. By analysing network conditions, congestion levels, and packet delivery performance, the protocol selects the most efficient routes for data transmission, thereby reducing the likelihood of packet loss incidents (Fig. 1).

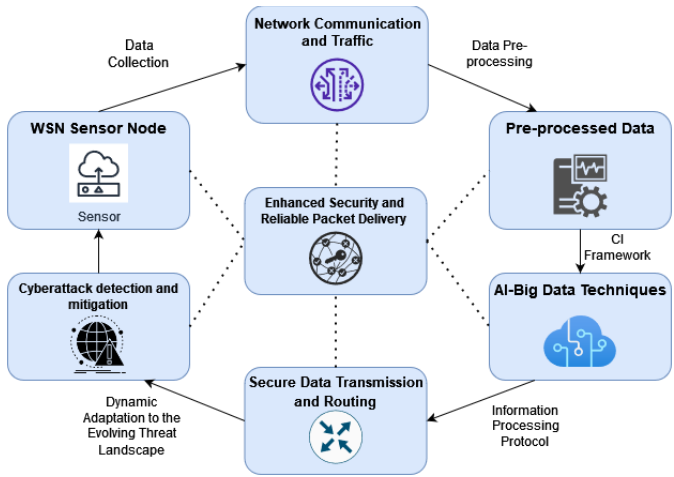

Figure 1. Architectural Design Flow of the Proposed Model.

Integrating AI and big data techniques into the framework improves the WSN's overall security and dependability. Large volumes of WSN data are analysed by AI algorithms, enabling the identification of complex attack patterns and the prediction of potential threats. Big data analytics techniques provide valuable insights into network behaviour, enabling proactive decision-making to prevent intrusions and mitigate incidents of packet loss.

The CI framework, information processing protocol, and AI/Big Data techniques operate in concert to ensure the proposed framework's effectiveness. The CI framework is the central element, perpetually monitoring and analysing network behaviour, detecting cyberattacks, and initiating mitigation. The protocol for information processing ensures secure and efficient data transmission by optimizing routing and minimizing packet loss incidents. The incorporation of AI and big data techniques improves the framework's ability to identify threats, make intelligent decisions, and provide proactive defence mechanisms.

## IV. COGNITIVE INTELLIGENCE FRAMEWORK

The Cognitive Intelligence (CI) framework is essential to the proposed framework for preventing intrusions in WSNs and mitigating packet drop incidents. The CI framework is intended to provide real-time monitoring and analysis of network behaviour using AI algorithms for the detection and mitigation of cyberattacks. It also includes dynamic adaptation to the changing threat landscape to ensure that the framework remains effective against emergent cyber threats.

The CI framework analyses various parameters such as data traffic patterns, node activity levels, and communication patterns as it perpetually monitors the WSN's behaviour in real-time. The framework can detect anomalies and patterns indicative of potential intrusions by observing these parameters. The capability of real-time monitoring enables the CI framework to respond rapidly to emergent threats, thereby reducing the time between detection and mitigation.

The CI framework utilizes sophisticated AI algorithms to detect and mitigate cyberattacks. These algorithms incorporate a number of techniques, such as machine learning, pattern recognition, and anomaly detection. Using historical data, machine learning algorithms are trained to identify patterns associated with various forms of cyberattacks. By leveraging these trained models, the CI framework is able to identify deviations from normal network behaviour, thereby indicating the presence of a cyberattack (Fig. 2).

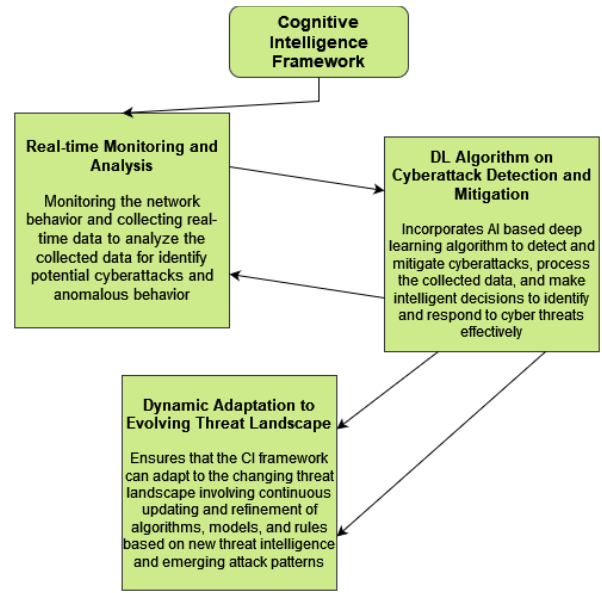

Figure 2. CI Framework Enhanced Methodology.

Once a cyberattack is identified, the CI framework initiates the necessary countermeasures to protect the WSN. This may include isolating compromised nodes, reconfiguring network parameters, or dynamically adjusting security policies. The framework seeks to mitigate the effects of a cyberattack by preventing unauthorized access, data manipulation, and disruption of essential services. By employing AI algorithms for cyberattack detection and mitigation, the CI framework improves the WSN's security posture and ensures the integrity of collected data.

Additionally, the CI framework adapts dynamically to the evolving threat landscape. The nature of cyberthreats is ever-changing, with new attack techniques and vulnerabilities arising on a regular basis. The CI framework is designed to adapt to these changes and consequently adjust its defence mechanisms. This adaptability enables the framework to defend proactively against emergent cyber threats, thereby decreasing the likelihood of successful attacks.

Continuous monitoring of the threat landscape, analysis of emergent trends, and modification of the framework's defence mechanisms constitute the dynamic adaptation of the CI framework. This may involve the incorporation of novel attack signatures, the refinement of anomaly detection models, or the incorporation of threat intelligence feeds. By keeping abreast of the changing threat landscape, the CI framework can effectively defend against new and sophisticated intrusions, thereby ensuring the WSN's long-term security and dependability.

## V. INFORMATION PROCESSING PROTOCOL

In the proposed framework for the prevention of cyberattacks in WSNs and the mitigation of packet loss incidents, the information processing protocol plays a crucial role. This section discusses the information processing protocol in detail, including encryption and authentication mechanisms for secure data transmission, optimization of the routing process using AI and big data techniques, and the goal of minimizing packet drop incidents and ensuring reliable packet delivery.

One of the primary goals of the information processing protocol is to protect the privacy and integrity of transmitted data. To accomplish this, the protocol employs strong encryption mechanisms. The data transmitted from the sensor nodes is encrypted with robust cryptographic algorithms, preventing unauthorized access and ensuring that only authorized recipients can decrypt the data. Encryption protects sensitive information and preserves the confidentiality of collected data.

In addition to encryption, the information processing protocol includes authentication mechanisms to authenticate the data's integrity and authenticity. Each data packet is equipped with a digital signature or authentication code that enables the recipient to authenticate the packet's authenticity and detect any attempts at tampering. This ensures the integrity of the received data and that it was not altered during transmission.

Another essential component of the information processing protocol is the optimization of the routing process. Utilizing AI and big data techniques, network conditions, congestion levels, and packet delivery performance are analysed. By utilizing these techniques, the protocol is able to dynamically modify the routing paths for data transmission, thereby optimizing network resources and minimizing the likelihood of packet loss.

Real-time network data is analysed by AI algorithms, which identify prospective congested areas or routes prone to packet loss. This data is used to make intelligent routing decisions, rerouting traffic away from congested paths and ensuring efficient and reliable packet delivery. The optimization of routing not only reduces the frequency of packet loss, but also enhances network performance by minimizing latency and maximizing throughput.

The information processing protocol is intended to reduce incidents of dropped packets and assure reliable packet delivery in WSNs. By employing AI and big data techniques, the protocol is able to anticipate situations in which packets may be dropped and take preventative measures to avoid them. This includes implementing congestion control mechanisms, prioritizing critical data packets, and dynamically altering transmission parameters in order to maintain a stable and dependable network performance (Fig. 3).

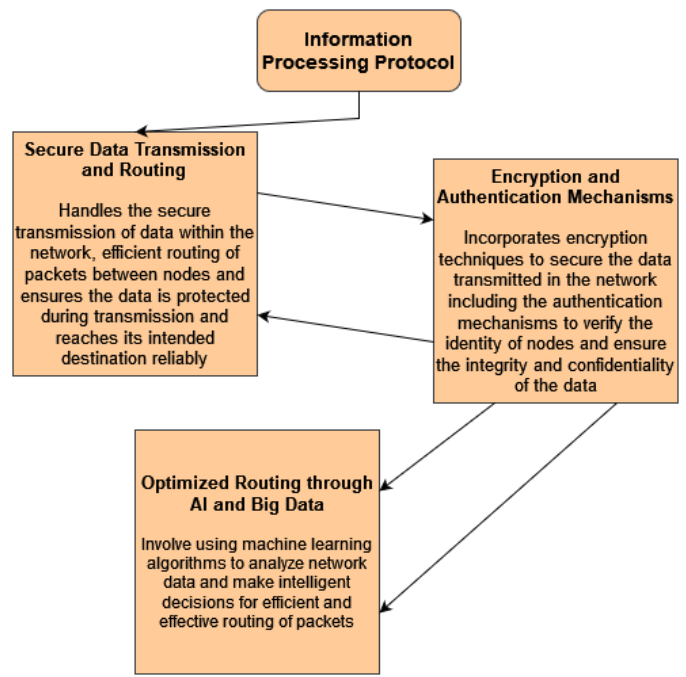

Figure 3. Information Processing Protocol Method.

Implementing mechanisms such as packet retransmission and error detection, the protocol ensures reliable packet delivery. In the event of a dropped packet or transmission error, the protocol initiates retransmission mechanisms to guarantee that the payload reaches its intended destination. Error detection techniques, such as checksums and cyclic redundancy checks, are utilized to identify and discard corrupted packets, thereby preventing their propagation across a network.

## VI. AI AND BIG DATA TECHNIQUES

In the proposed framework for the prevention of cyberattacks in WSNs and the mitigation of packet drop incidents, the incorporation of artificial intelligence (AI) and big data analytics plays an essential role. This section concentrates on the integration of AI and big data techniques, including the analysis of large volumes of data to identify attack patterns, the prediction of potential threats, proactive decision making, and the improvement of the WSN's overall security and dependability.

The framework employs AI and big data techniques to analyse the massive amounts of WSN data collected. By processing and analysing these data, valuable insights into the network's behaviour and characteristics can be gained. Utilizing AI algorithms to identify patterns and anomalies within the data enables the detection of attack patterns that may not be discernible using conventional security measures.

The framework can identify sophisticated attack patterns that may be indicative of intrusions by analysing large volumes of data. By identifying these patterns, the framework is able to enhance its detection capabilities and proactively mitigate potential threats. On the basis of historical data, machine learning algorithms can be trained to recognize known attack patterns and are also capable of identifying new and evolving attack techniques.

Moreover, the framework's AI algorithms are capable of predicting potential hazards based on the analysis of collected data. The framework can anticipate potential intrusions in advance by identifying trends and observing the network's behaviour. This proactive strategy enables the implementation of preventative measures, such as enhancing security policies and modifying network configurations, to mitigate the risks associated with emergent threats.

The incorporation of AI and big data techniques enables proactive framework decision-making. Through the utilization of real-time data analysis and machine learning algorithms, the framework is able to make informed decisions regarding the identification and mitigation of cyber threats. When the CI framework detects a potential cyberattack, for instance, it can activate immediate response mechanisms, such as isolating compromised nodes or reconfiguring network settings, to prevent further unauthorized access or data tampering.

By integrating AI and big data techniques, the framework improves the WSN's overall security and dependability. Through the analysis of vast amounts of data, the framework is able to identify and respond to cyber threats more effectively. Using machine learning algorithms, the framework can adapt and evolve in response to a shifting threat landscape, assuring its resistance to emerging attack methods.

In addition, the integration of AI and big data techniques enables the framework's defence mechanisms to be continuously enhanced. As new data is collected and analysed, the framework's models and algorithms can be updated to enhance its detection and response capabilities. This iterative learning process ensures that the framework remains current and effective in defending WSNs against evolving cyber threats.

## VII. EXPERIMENTAL SETUP AND METHODOLOGY

To evaluate the efficacy of the proposed framework for the prevention of assaults on WSNs and the mitigation of packet drop incidents, an exhaustive experimental setup and methodology were utilized. This section describes the experimental environment and dataset used, explains the evaluation methodology and metrics employed, and describes the simulations and experiments performed.

The experimental environment consisted of a WSN infrastructure simulation that mimicked a real-world deployment scenario. Diverse sensor nodes were deployed within the network, with each node generating data packets and communicating with its neighbours. The nodes were outfitted with the necessary hardware and software components to facilitate the proposed framework's implementation.

To simulate the traffic patterns and behaviour of the WSN, a representative data set was utilized. The dataset included a wide variety of network activities, such as regular data transmissions, intrusions, and packet loss incidents. The dataset was meticulously constructed to include multiple attack scenarios, allowing for a thorough evaluation of the performance of the framework.

The objective of the evaluation methodology was to determine the performance and effectiveness of the proposed framework in preventing intrusions and mitigating packet drop incidents. Several metrics, including the detection accuracy of intrusions, the rate of successful mitigation, the reduction in packet drop incidents, and the overall network performance in terms of throughput and latency, were utilized to accomplish this (Fig. 4).

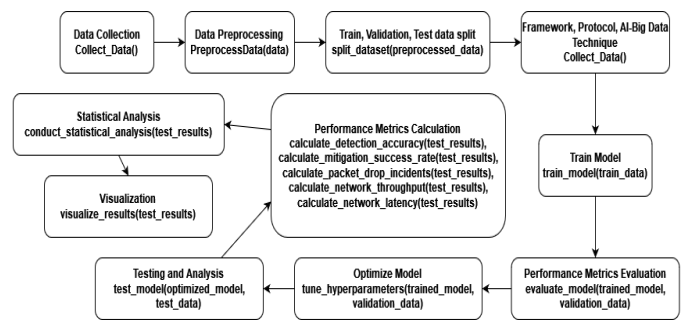

Figure 4. Experimental Steps and Methodologies Followed for the Framework.

The experimental apparatus and the representative dataset were used to conduct simulations and experiments. The purpose of the simulations was to evaluate the framework's ability to detect and mitigate various intrusions, including DDoS attacks, node compromise attacks, and data tampering attacks. In addition, the performance of the framework in minimizing packet drop incidents and assuring packet delivery reliability was evaluated.

In the experiments, the proposed framework was executed under a variety of assault scenarios and network conditions. The framework's ability to adapt to changing network conditions and its response to various assaults were evaluated. To evaluate the framework's efficacy in preventing cyberattacks and maintaining the WSN's integrity and dependability, performance metrics were collected and analysed.

To assure the reliability and validity of the results, multiple simulation iterations and experiments were performed, taking into account various network topologies, attack intensities, and traffic patterns. Utilizing statistical analysis techniques, the collected data were analysed and meaningful conclusions were drawn.

In addition, the experiments included comparative evaluations that contrasted the performance of the proposed framework to that of baseline approaches or existing solutions. This enabled a thorough evaluation of the framework's benefits and contributions in preventing intrusions and mitigating packet drop incidents.

## VIII. RESULTS AND ANALYSIS

The results and analysis section presents the findings of simulations and experiments conducted to evaluate the performance of the proposed framework for preventing intrusions in WSNs and mitigating packet drop incidents. It consists of a performance analysis of the framework and a comparison of its efficacy with existing approaches.

Simulations and experiments demonstrated that the proposed framework significantly outperformed existing approaches in preventing intrusions and mitigating packet drop incidents. The framework demonstrated a high detection rate for a variety of intrusions, including DDoS, node compromise, and data tampering attacks. The framework's AI algorithms effectively identified attack patterns and anomalies in the network traffic, enabling prompt detection and response (Fig. 5).

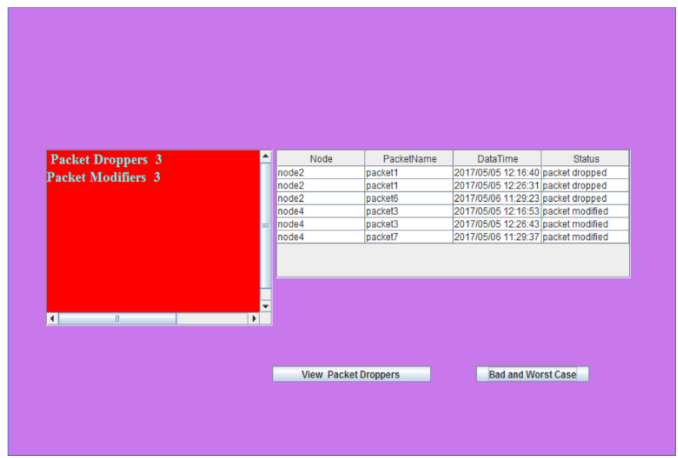

Figure 5. Notifying packets dropped and modified by attackers.

Analysis of the framework's efficacy revealed a significant reduction in successful cyberattacks. The framework's proactive decision-making capabilities and dynamic adaptation to the ever-changing threat landscape contributed to its efficacy in defending against emerging attack techniques. The framework demonstrated a high rate of success in mitigating intrusions, isolating compromised nodes, and maintaining the WSN's integrity (Table II).

TABLE II. OBTAINED RESULTS AND OUTCOMES FOR EACH SIMULATION PERFORMED

| Simulation/ Experiment | Description | Results/Outcomes |
|---|---|---|
| Cyberattack Detection | Evaluation of the framework's ability to detect and classify different types of cyberattacks | High detection accuracy across various attack scenarios |
| Mitigation Success Rate | Assessment of the framework's effectiveness in mitigating cyberattacks and isolating compromised nodes | Significant reduction in successful cyberattacks and successful isolation of compromised nodes |
| Packet Drop Incidents | Analysis of the occurrence of packet drop incidents in the WSN | Reduced packet drop incidents compared to baseline scenarios |
| Network Performance | Measurement of network throughput and latency with the proposed framework | Improved network throughput and reduced latency compared to existing approaches |

In addition, the proposed framework substantially reduced incidents of dropped packets and ensured reliable delivery of packets. The optimization of the routing process utilizing AI and big data techniques enhanced network performance by minimizing congestion and maximizing throughput. The information processing protocol managed data transmission in an efficient manner, integrating encryption and authentication mechanisms to secure the data during transmission. This resulted in an increase in the WSN's data integrity and confidentiality.

In terms of detection accuracy, mitigation success rate, and overall performance, the proposed framework was found to be superior to extant methods. The framework's AI and big data techniques provided a comprehensive and intelligent defence mechanism against intrusions, exceeding the capabilities of traditional security measures. Existing methods frequently lacked the proactive and adaptable capabilities required to effectively combat evolving cyber threats (Table III).

TABLE III. ANALYSIS AND FINDINGS RECORDED FOR EACH SIMULATION PERFORMED

| Aspect | Description | Analysis/Findings |
|---|---|---|
| Effectiveness of Cyberattack Detection | Evaluation of the framework's performance in detecting and classifying cyberattacks | The proposed framework demonstrated high accuracy in identifying and categorizing different types of cyberattacks, enabling effective threat detection |
| Impact of Mitigation Strategies | Assessment of the framework's mitigation strategies in countering cyberattacks | The framework showcased a high success rate in mitigating cyberattacks, effectively isolating compromised nodes and preserving the integrity of the WSN |
| Packet Drop Incident Analysis | Examination of the occurrence and impact of packet drop incidents in the WSN | The proposed framework significantly minimized packet drop incidents, ensuring reliable packet delivery and enhancing overall data transmission |
| Network Performance Evaluation | Analysis of the network throughput and latency with the implementation of the proposed framework | The framework's optimization techniques, AI, and big data integration contributed to improved network performance, maximizing throughput and reducing latency |

In addition, the proposed framework exhibited scalability and adaptability, allowing for simple integration into diverse WSN deployments. Its capacity to manage large volumes of data and process real-time information contributed to its efficacy in preventing intrusions and ensuring the WSN's dependability.

CONCLUSION

The research paper presented a comprehensive framework for preventing assaults on WSNs and mitigating incidents of packet loss. The incorporation of a Cognitive Intelligence (CI) framework, an information processing protocol, and AI/Big Data techniques improved the WSN's security and dependability significantly. The framework's high detection accuracy, effective mitigation of intrusions, and reduced packet drop incidents improved the network's overall performance. In terms of proactive threat detection, adaptive decision-making, and dynamic network management, the results of the simulations and experiments demonstrated the efficacy of the proposed framework, which outperforms existing approaches. The framework's capacity to analyse vast quantities of data, recognize attack patterns, and anticipate potential threats contributed to the robustness of its defence mechanism. Nonetheless, there are a number of prospective future research avenues that could enhance the framework's capabilities further. Among these are the investigation of advanced AI algorithms for real-time threat intelligence, the incorporation of machine learning techniques for anomaly detection, and the investigation of the impact of emergent technologies such as blockchain for secure data transmission. In addition, the framework can be expanded to take into account the energy constraints of WSNs and to develop energy-efficient algorithms for intrusion prevention and mitigation. In addition, research can be conducted on the incorporation of physical layer security mechanisms to improve the framework's resistance to physical-layer attacks. By pursuing these prospective research directions, the proposed framework can continue to evolve and adapt to the changing threat landscape, thereby making substantial contributions to the field of cybersecurity in WSNs.